\documentclass[10pt,notitlepage]{article}

\usepackage{graphicx}
\usepackage{amsmath}
\usepackage{amssymb}
\usepackage{algorithm}
\usepackage{algorithmic}
\usepackage{hyperref}
\usepackage{url}
\usepackage{booktabs}
\usepackage{xcolor}

\usepackage[margin=1in]{geometry}

\date{}

\title{Dynamic Evaluation Framework for Personalized and Trustworthy Agents: \\A Multi-Session Approach to Preference Adaptability}

\title{Dynamic Evaluation Framework for Personalized and Trustworthy Agents: A Multi-Session Approach to Preference Adaptability\\
\large{}\vspace*{2em}\textsuperscript{*}Chirag Shah (chirags@uw.edu), \textsuperscript{†}Hideo Joho (joho.hideo.gb@u.tsukuba.ac.jp),\\
\textsuperscript{*}Kirandeep Kaur (Kaur13@cs.washington.edu), \textsuperscript{*}Preetam Prabhu Srikar Dammu (preetams@uw.edu)\\
\small{}\textsuperscript{*}University of Washington, Seattle, USA\\
\textsuperscript{†}University of Tsukuba, Tsukuba, Japan
}

\begin{document}

\maketitle

\begin{abstract}
Recent advancements in generative AI have significantly increased interest in personalized agents. With increased personalization, there is also a greater need for being able to trust decision-making and action taking capabilities of these agents. However, the evaluation methods for these agents remain outdated and inadequate, often failing to capture the dynamic and evolving nature of user interactions. In this conceptual article, we argue for a paradigm shift in evaluating personalized and adaptive agents. We propose a comprehensive novel framework that models user personas with unique attributes and preferences. In this framework, agents interact with these simulated users through structured interviews to gather their preferences and offer customized recommendations. These recommendations are then assessed dynamically using simulations driven by Large Language Models (LLMs), enabling an adaptive and iterative evaluation process. Our flexible framework is designed to support a variety of agents and applications, ensuring a comprehensive and versatile evaluation of recommendation strategies that focus on proactive, personalized, and trustworthy aspects.
\end{abstract}

\section{Introduction}
The rise of AI-driven personalized agents has transformed how users interact with information, from recommendation systems to conversational agents. These agents must effectively infer and adapt to users' evolving needs, requirements, and preferences to provide meaningful personalization. While such agents have received a lot of attention lately, their future depends on our ability to evaluate them reliably. Evaluating them using the traditional metrics for a typical Information Retrieval (IR) systems is going to be insufficient at best and misleading at worst. We need a novel way to think about what such personalized agents are meant to do and how we should evaluate their effectiveness. This opens up a new area of research in IR that is rooted in the existing literature, but has nuances that are unexplored. For example, traditional evaluation methods often rely on static datasets or predefined user profiles, limiting their ability to assess an agent's adaptability in real-world interactions. This limitation hinders the development of truly adaptive agents that can meet users' dynamic needs in real-time. 

In this conceptual paper, we envision a novel evaluation framework that leverages simulated user personas and structured interactions to assess personalized agents. Inspired by techniques from reference interviews in information retrieval~\cite{dervin1986neutral}, our proposal involves agents conducting preference elicitation dialogues with simulated users. These interactions allow agents to iteratively refine their understanding of user needs, making the evaluation process more adaptive and user-centric.  

Our framework consists of three core components: (1) Simulated User Personas with distinct and evolving preferences; (2) Structured User-Agent Interactions, where agents extract preferences and provide recommendations; and (3) Iterative Feedback Mechanisms, enabling agents to refine their personalization strategies over multiple interactions. Unlike traditional approaches, which evaluate recommendations against static benchmarks, our framework incorporates real-time feedback mechanism via an LLM-based simulation, allowing a more nuanced assessment of agent performance. 

Our proposal addresses the critical gap in evaluating the adaptability of personalized agents by posing the question: \emph{How can we assess an agent's ability to dynamically model user preferences in real-time?}. Traditional evaluation methods, which often rely on static datasets and predefined user profiles, are inadequate for assessing how well agents can adapt to changing user needs and preferences over time. To bridge this gap, we adopt a theory-driven approach to systematically understand the evolution of personalization in information-seeking tasks and propose our framework. We begin by presenting a systematic evolution of personalization in information retrieval tasks in Section~\ref{sec:personalzationevolution}. Drawing insights from the literature, we define the concept of dynamic evaluation for personalized agents and outline the core components involved in the evaluation of such agents in Section~\ref{sec:dynamicevaluation}. We then delve into these components to propose our framework in Section~\ref{sec:framework} with a detailed cases study example of travel planning. While we conceptually demonstrate our framework on travel planning, it is designed to be highly adaptable and generalizable, making it equally effective for a range of domains, including e-commerce, entertainment, and online shopping. By introducing a structured and interactive evaluation methodology, we aim to contribute to the development of more intelligent and responsive personalized AI and advance the field of adaptive user modeling~\cite{ijcai2019p585}, enabling robust evaluation of personalized agents that can dynamically respond to user preferences across diverse contexts.

\section{Personalization} 
\label{sec:personalzationevolution} 
Traditional IR systems, based on keyword matching and Boolean operators, often led to information overload and irrelevant results~\cite{carlson2003information}. These systems lacked contextual understanding, resulting in inefficiencies. The advent of personalization models marked a significant shift by automating search processes tailored to individual needs using historical interactions and contextual information~\cite{Harman1992}. The core idea is to improve relevance by aligning results with user interests. Techniques such as \emph{user profiling} create profiles based on past queries and interactions~\cite{Bennett2007}, enabling systems to anticipate interests and refine search outcomes. \emph{Collaborative filtering} recommends items based on the preferences of similar users~\cite{Koren2009}, used by platforms like Netflix~\cite{Ricci2015}. \emph{Content-based filtering} focuses on item characteristics, recommending similar content~\cite{Sarwar2001}, but tends to over-specialize~\cite{pazzani1999framework}. Personalized agents also use \emph{contextual information} to refine personalization by considering factors like time of day, location, or device type, making real-time adjustments~\cite{Lops2011}. Despite these improvements, traditional IR systems are limited by their static nature~\cite{sloan2015dynamic}. The emergence of generative personalized agents marks a paradigm shift, introducing dynamic, real-time adaptation to user preferences.

Traditional personalization approaches are static and reactive, adapting only when new data is explicitly provided. In contrast, LLM-based personalized assistants generate real-time responses, dynamically adjusting to user preferences, contextual cues, and conversational histories~\cite{radlinski2017theoretical,10.1145/3477495.3532678}. Recent advancements highlight LLMs' potential in personalizing information retrieval, contextualizing user queries~\cite{tan2024efficient,kaur2025efficient}, augmenting retrieval through effective query generation~\cite{baek2023knowledge}, and adapting responses to evolving preferences~\cite{salemi2023personalized}. Multimodal data integration makes interactions more immersive~\cite{salemi2024optimizing}, and efficient fine-tuning methods improve alignment with user-specific needs~\cite{dl2024grounded}. Persona-based agents use explicit personas for nuanced understanding. The \textit{Persona-DB} framework distills personas from interaction histories for accurate context-aware responses~\cite{sun2024persona}. The \textit{PICLe} framework employs Bayesian inference for diverse persona elicitation, enhancing conversational richness~\cite{choi2024picle}. Persona-driven conversational datasets improve response consistency and relevance~\cite{findings2024faithful}. Persona-based LLMs enhance AI systems' steerability, aligning responses with specific personas~\cite{amazon2024steerability}. Techniques like open-ended life narratives condition LLMs to adopt distinct virtual personas, improving response consistency and fairness~\cite{anthology2024}. PROXONA applies LLM-driven personas in content creation, refining content strategies dynamically~\cite{proxona2024}.

Adaptive user modeling began with sequential and reinforcement learning techniques, dynamically adjusting recommendations based on evolving behaviors~\cite{shani2005mdp,Zhang2017DeepLB}. These approaches used Markov Decision Processes (MDPs) and recurrent neural networks (RNNs) to capture short-term and long-term interests. In generative dialogue systems, adaptive generation techniques enhance conversational agents, ensuring responses evolve with changing preferences~\cite{janarthanam2014dialogue}. The Cognitive Personalized Search (CoPS) model integrates large language models with a cognitive memory structure inspired by human cognition, enhancing user modeling and search experiences, particularly in zero-shot scenarios~\cite{li2024adaretrievaladaptivemultiroundretrieval}. Despite advancements, current models lack comprehensive evaluation frameworks for assessing true adaptation to changing behaviors. Most rely on static datasets or synthetic interactions, failing to capture evolving needs in real-world settings. Metrics like response relevance, engagement scores, or next-query prediction accuracy provide limited insights into long-term adaptation. Reinforcement learning-based approaches optimize for short-term objectives but struggle with delayed feedback and long-term preference drifts~\cite{shani2005mdp}. Retrieval-augmented generation (RAG) models and dynamic memory mechanisms refine user representations at a single-session level but do not evaluate sustained personalization effectiveness across multiple interactions. A robust evaluation framework is needed -- one that incorporates real-time feedback, longitudinal adaptation studies, and task-oriented assessments reflecting true evolution in preferences. Without such frameworks, adaptive models risk being optimized for static metrics rather than true user-centric adaptability.

\section{Dynamic Evaluation}
\label{sec:dynamicevaluation}
In evaluating personalized information retrieval (IR) models, traditional metrics like precision, recall, mean reciprocal rank (MRR), mean average precision (MAP), and normalized discounted cumulative gain (NDCG) are commonly used to assess system performance. These metrics focus on the accuracy of retrieval at a specific point in time, evaluating how well the system retrieves relevant information based on static queries and user profiles~\cite{bai2018ecnu}. However, they often fail to capture the dynamic nature of user preferences and behaviors as they evolve over time~\cite{tabrizi2018person}. Recent works have developed evaluation frameworks that incorporate such metrics, user profiles, and interaction histories to capture a system's adaptability to changing user preferences~\cite{tabrizi2018person}.

The emergence of LLM-based personalized agents introduces greater challenges for evaluation. Unlike traditional IR systems, which operate on session-based or short-term interactions, LLM-based agents maintain prolonged engagement with users, often spanning months or years. These agents continuously adapt to users' evolving preferences, goals, and contextual needs, requiring evaluation frameworks that go beyond static or short-term personalization~\cite{radlinski2017theoretical}. Existing methodologies struggle to measure whether an LLM-based agent effectively tracks long-term user preference shifts, refines its recommendations over time, or mitigates challenges such as preference drift and forgetting. Moreover, conventional personalization metrics, which depend on predefined relevance judgments, fail to accommodate the open-ended and generative nature of LLMs, where responses dynamically evolve based on user interactions.

Thus, there is a pressing need for evaluation frameworks that can assess the effectiveness of LLM-based personalized agents in long-term user adaptation. These frameworks must extend beyond static relevance-based evaluation, incorporating dynamic and longitudinal assessments to determine whether an agent effectively retains, updates, and refines user preferences while maintaining robustness. To bridge this gap, we propose key components for assessing dynamic personalized agents:

\begin{itemize}
    \item \textbf{Simulated User (Persona/SIM)}: Enables controlled experiments on how systems respond to evolving user behaviors, acting as a virtual persona with predefined attributes and preferences that evolve over successive interactions.
    \item \textbf{Personalized Agent}: Engages in continuous interactions with the simulated user, focusing on updating and refining recommendations based on evolving user profiles.
    \item \textbf{Task-Based Interaction (Work Task)}: Anchors evaluations in realistic tasks to assess system performance within practical, goal-oriented contexts~\cite{Braggaar2023EvaluatingTD}.
    \item \textbf{Personalization Elicitation (Reference Interview)}: Uses structured interviews and think-aloud protocols to extract and iteratively refine user preferences, mirroring real-world interactions~\cite{ericsson1984protocol}.
    \item \textbf{Dataset}: Provides a structured collection of items with associated attributes, serving as the foundation for the agent's retrieval and ranking process.
    \item \textbf{Ranked Items}: Reflects how well the agent adapts to evolving user needs over time by retrieving and ranking items from the dataset based on inferred preferences.
    \item \textbf{Dynamic Evaluation and Measurements}: Ensures the agent's personalization effectiveness is assessed iteratively, capturing how recommendations improve as user interactions progress.
\end{itemize}

\begin{figure}
    \centering
    \includegraphics[width=\linewidth]{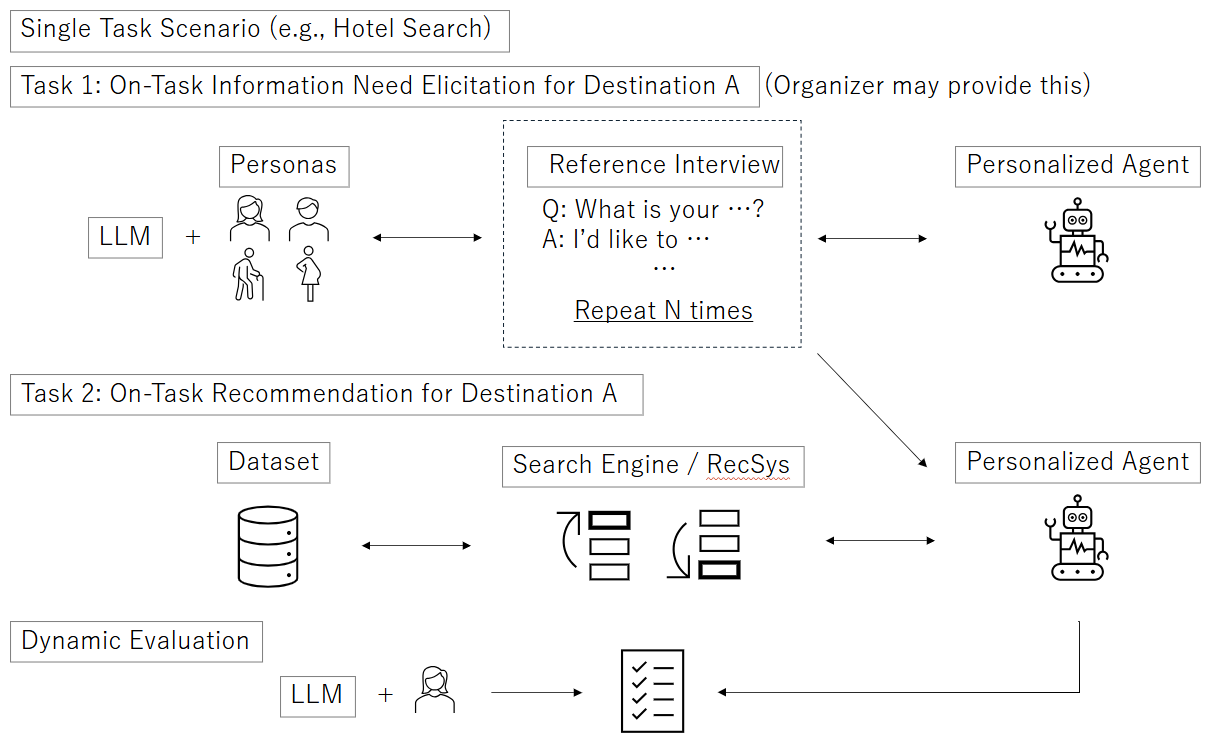}
    \caption{Overview of our evaluation framework showing the core components and their interactions.}
    \label{fig:framework}
\end{figure}

\section{Evaluation Framework}
\label{sec:framework}
An overview of our evaluation framework is shown in Figure \ref{fig:framework}. As can be seen, our framework consists of seven major components: Persona/SIM, Personalized Agent, Tasks, Datasets, Ranked Items, Dynamic Evaluation and Measurements.

\subsection{Work Task: Travel Planning}
This section leverages aforementioned key components to operationalize our evaluation framework using the travel planning as the work task of end users. To demonstrate the utility of our framework, we further define three sub work tasks such as \textit{Flight finding}, \textit{hotel finding}, and \textit{restaurant finding}. Furthermore, we assume that there are two destinations for travel planning: \texttt{Destination A} and \texttt{Destination B}, where the travel planning of \texttt{Destination B} occurs \textit{after} that of \texttt{Destination A}. This allows us to evaluate the agent's performance of using the previous interaction data for a new but similar type of task. Specifically, we examine how effectively the agent:
\begin{itemize}
    \item Adapts to user preferences based on prior interactions,
    \item Refines recommendations over time, and
    \item Balances exploration vs. exploitation in various sub tasks to optimize user satisfaction.
\end{itemize}

\subsection{Step 1: Reference Interview}

The first step in our evaluation framework assesses the agent's ability to elicit user needs, requirements, and preferences for the given task through a reference interview. This interaction is crucial for tailoring recommendations, as it allows the agent to gather key details about the user's intent. While the actual impact of this interview will be reflected in the ranking and selection of recommended options in subsequent steps, the agent's ability to engage in effective, dynamic, and context-aware question-answering conversations is a fundamental competency.

It is important to note that the reference interview differs from conventional conversational sequential recommendations. In conversational recommendation, it assumes that a user feedback is available for each of recommended items and update the user profile after each feedback. In many real-world scenario, the user is not available in such a manner. More often, an agent has a limited amount of time to gather information from the client once, and conduct a research based on the information and make recommendations. We adapt the latter case because:
\begin{itemize}
 \item It allows us to evaluate different aspects of personalized agents beyond conventional interactive recommendation systems.
 \item It increases the reproducibility of experimental findings.
 \item It gives researchers the flexibility to focus on different technical challenges, such as elicitation strategies, preference modeling, or adaptive response generation.
\end{itemize}

\subsection{Step 2: Item Recommendation}
The second step in our evaluation framework assesses the agent's ability to retrieve and rank items from the dataset based on the needs, requirements, and preferences elicited in Step 1. This step encompasses standard recommendation and ranking techniques, but our framework extends beyond conventional approaches to address key challenges in personalized shopping assistance.
Unlike traditional recommendation systems that rely on static user profiles or collaborative filtering, our evaluation focuses on dynamic personalization, measuring how well the agent interprets the elicited needs and requirements to generate relevant and diverse product recommendations; balances personalization and diversity by incorporating both user preferences and exploratory suggestions; adapts recommendations over multiple sessions, leveraging prior interactions from Session A to refine recommendations in Session B. Thus, this step includes a standard recommendation/ranking element, but our framework goes beyond to cover advanced challenges as follows.

\subsubsection{Challenge 1: On-Task Item Recommendation for \texttt{Session A}}
This challenge takes as input the shopping task description, sub-task description (e.g., hotel finding), dataset description, and QA utterances from Step 1, and returns a ranked list of products that align with the Needs, Requirements, and Preferences of users. This assesses that the agent effectively interprets user intent and ranks products accordingly. The retrieved and ranked list should reflect user-stated preferences while maintaining a balance between relevance, diversity, and novelty.

\subsubsection{Challenge 2: On-Task Item Recommendation for \texttt{Session B}}
This challenge takes the same set of descriptions as Challenge 1, but the personal agent is asked to recommend items for \texttt{Destination B}. Inputs can include the user feedback on the recommended items for \texttt{Destination A}. This challenge evaluates whether the personal agent can infer user's Needs, Requirements, and Preferences (NRP) from the previous session on the same sub task to a different destination. For example, if a user selected a particular brand of hotel in \texttt{Destination A}, the same brand might be relevant in \texttt{Destination B}. 

\subsubsection{Challenge 3: Cross-Task Item Recommendation for \texttt{Session A}}
This challenge takes the same set of descriptions as Challenge 1, but the personal agent is asked to recommend items for another sub-task (e.g., flight finding). Inputs can include the user feedback on the recommended items on other sub tasks (e.g., hotel finding). This challenge evaluates whether the personal agent can infer user's NRP from the previous session on the same destination to a different but related sub task. For example, the dietary requirement identified by the restaurant finding subtask can be used to determine relevant items in hotel finding subtask.

\begin{algorithm}
\caption{Cross-Task Item Recommendation}
\label{alg:cross_task_recommendation}
\begin{algorithmic}
\STATE \textbf{Input:} Task description $T$, Sub-task description $S_1$, Dataset description $D$, QA utterances $U$, User feedback on $S_1$ recommendations $F$, New sub-task description $S_2$
\STATE \textbf{Output:} Ranked list of items for $S_2$
\STATE $P \gets $ ExtractPreferences($U$, $F$)
\STATE $P_{cross} \gets $ InferCrossTaskPreferences($P$, $S_1$, $S_2$)
\STATE $R \gets $ RetrieveItems($D$, $S_2$, $P_{cross}$)
\STATE $R_{ranked} \gets $ RankItems($R$, $P_{cross}$)
\RETURN $R_{ranked}$
\end{algorithmic}
\end{algorithm}

\subsubsection{Challenge 4: Cross-Task Item Recommendation for \texttt{Session B}}
This challenge evaluates the agent's ability to transfer knowledge across both tasks and sessions. For example, if the user preferred boutique hotels with high cleanliness ratings in \texttt{Destination A}, the agent might recommend boutique car rental services with personalized customer support in \texttt{Destination B}. This ensures the agent can infer and apply user preferences from multiple tasks in Session A to a different task in Session B, enhancing the overall user experience.

\begin{algorithm}
\caption{Cross-Task Cross-Session Retrieval and Ranking}
\label{alg:cross_task_retrieval_ranking}
\begin{algorithmic}
\STATE \textbf{Input:} Task descriptions for Session A $T_A$ and Session B $T_B$, 
\STATE Sub-task descriptions $S_{A1}$, $S_{A2}$, $S_{B1}$, QA utterances $U_A$, $U_B$,
\STATE User feedback on Session A recommendations $F_A$
\STATE \textbf{Output:} Ranked list of items for new sub-task in Session B $S_{B2}$
\STATE $P_A \gets $ ExtractPreferences($U_A$, $F_A$, $S_{A1}$, $S_{A2}$)
\STATE $P_B \gets $ ExtractPreferences($U_B$, $S_{B1}$)
\STATE $P_{combined} \gets $ CombinePreferences($P_A$, $P_B$, $T_A$, $T_B$)
\STATE $P_{cross} \gets $ InferCrossTaskPreferences($P_{combined}$, $S_{B1}$, $S_{B2}$)
\STATE $R \gets $ RetrieveItems($D$, $S_{B2}$, $P_{cross}$)
\STATE $R_{ranked} \gets $ RankItems($R$, $P_{cross}$)
\RETURN $R_{ranked}$
\end{algorithmic}
\end{algorithm}

\subsection{Step 3: Dynamic Evaluation}
Evaluating personalized agents require a robust and adaptable framework that can assess relevance across diverse user needs and dynamic interaction contexts. Our framework introduces dynamic evaluation through a simulated LLM-based user, which systematically labels recommendations based on their alignment with user preferences. This allows for continuous benchmarking of different recommendation strategies, ensuring that models generalize well beyond static datasets.  

As illustrated in Figure \ref{fig:framework}, the simulated user assesses the match between the attributes of recommended items and the attributes of its persona (SIM). This process enables an automated, scalable, and repeatable evaluation of recommendation quality, making it possible to test new models and algorithms under controlled but realistic conditions. Unlike traditional static evaluations, this approach facilitates a deeper understanding of how recommendation agents adapt to user preferences over time, especially in multi-session and cross-task scenarios.  

To capture the full spectrum of recommendation performance, our framework supports a diverse set of evaluation metrics tailored to the PersonalWAB dataset. First, relevance-based metrics such as $NDCG@K$ and $Precision@K$ measure the quality and ranking of recommendations, ensuring that only highly relevant items are surfaced to users. Next, personalization metrics, including Personalization Score and Coverage, assess how well the agent differentiates between users and explores the full item space. The framework also includes adaptability metrics such as Cross-Session Consistency and Novelty Score, which evaluate whether the agent infers long-term user preferences and introduces diverse but relevant recommendations over multiple interactions. Lastly, trustworthiness and robustness metrics, including Fairness Evaluation and Robustness to Noisy Preferences, ensure that recommendations remain unbiased and stable even under ambiguous or conflicting user signals.  

Our framework is deeply aligned with the PersonalWAB dataset, which provides a rich benchmark for evaluating personalized agents in multi-session, multi-task, and real-world scenarios. It enables the testing of on-task recommendation challenges (e.g., suggesting products in a given category) as well as cross-task recommendation challenges (e.g., inferring shopping preferences from previously purchased categories). The synthetic user feedback loops within PersonalWAB ensure that evaluations reflect realistic user interactions, allowing researchers to test whether recommendation agents improve over time based on past user feedback.  

By integrating dynamic simulation, personalized evaluation metrics, and real-world task complexity, our framework establishes a comprehensive, scalable, and reproducible methodology for benchmarking personalized recommendation agents. This approach not only enhances the interpretibility of model performance but also provides actionable insights into designing more adaptive and user-centric recommendation systems.  

\section{Conclusion and Future Work}
In this paper, we introduced and validated a novel evaluation framework designed to assess the adaptability and effectiveness of AI-driven personalized agents. Our framework addresses the limitations of traditional evaluation methods by incorporating dynamic user profiles and real-time interaction data, thereby providing a more comprehensive assessment of an agent's ability to meet users' evolving Needs, Requirements, and Preferences (NRP).

The validation of our framework demonstrated its potential to significantly enhance the development of personalized agents, ensuring they can adapt to users' dynamic needs in real-world scenarios. This advancement paves the way for future research and development in the field of personalized AI.

Looking ahead, several research directions and questions emerge from this proposed framework:

\begin{itemize}
    \item {\bf Longitudinal Studies}: How do personalized agents perform over extended periods, and how do they adapt to long-term changes in user behavior and preferences?
    \item {\bf Cross-Domain Adaptability}: Can personalized agents effectively transfer learned preferences and behaviors across different domains or contexts?
    \item {\bf User Privacy and Ethical Considerations}: How can we ensure that personalized agents respect user privacy and adhere to ethical standards while collecting and utilizing real-time interaction data?
    \item {\bf Scalability}: How does the proposed framework scale with increasing numbers of users and more complex interaction scenarios?
    \item {\bf Human-AI Collaboration}: What are the best practices for designing personalized agents that can seamlessly collaborate with humans, enhancing productivity and user satisfaction?
\end{itemize}

By exploring these questions, we can further refine and expand the capabilities of personalized agents, ultimately creating more adaptive, responsive, trustworthy, and user-centric AI systems. Our proposed framework serves as a foundational step towards achieving this vision, fostering innovation and progress in the realm of AI-driven personalization.

\clearpage
\bibliographystyle{plain}
\bibliography{references}

\end{document}